\documentclass[pre,twocolumn,showpacs,amssymb,floatfix]{revtex4}
\usepackage{graphicx}
\usepackage[breaklinks]{hyperref}
\usepackage[tight]{subfigure}
\usepackage{amsmath}

\def\Mout{M_\mathrm{out}}

\begin{document}

\title{Evaluating Local Community Methods in Networks}

\author{James P.~Bagrow}
\affiliation{Department of Physics, Clarkson University, Potsdam NY 13699-5820 USA}
\email{bagrowjp@clarkson.edu}

\date{October 4, 2007} % ***

\begin{abstract}  
We present a new benchmarking procedure that is unambiguous and specific to local community-finding methods, allowing one to compare the accuracy of various methods.  We apply this to new and existing algorithms.  A simple class of synthetic benchmark networks is also developed, capable of testing properties specific to these local methods.  
\end{abstract}

% 89.75.Da,	% Systems obeying scaling laws
% 02.50.Ey  % Stochastic processes

\maketitle

\section{Introduction}
The study of complex networks~\cite{strogatzReview:Exploring:2001,citeulike:556495,newman:networks_review2003} has recently arisen as a powerful tool for understanding a variety of systems, such as biological and social interactions~\cite{WattsStrogatzSmallWorldNature,BA:nature:metabolic}, technology communications and interdependencies~\cite{strogatzReview:Exploring:2001,faloutsoi-topologies}, and many others.  The problem of detecting \emph{communities}, subsets of network nodes that are densely connected amongst themselves while being sparsely connected to other nodes, has attracted a great deal of interest due to a variety of applications~\cite{newmangirvan:communities,newmangirvan:mixingpatterns:book:2003,newman:2004:epj, rccvp_defining,portermuchanewman:uscongress,newman:pnas:review:2006}. Many techniques have been developed to find these subsets, with a broad array of costs and associated accuracies~\cite{citeulike:332503}.

Many community-finding algorithms hinge upon maximizing a quantity known as Modularity~\cite{newman:2004:pre,newmanclausetmoor:fastmodularitypre:2004}, often defined as:
\begin{equation}
	Q = \frac{1}{2M} \sum_{v,w}\left(A_{vw} - \frac{k_v k_w}{2M}  \right) \delta(c_v,c_w),
	\label{eqn:modularityDefined}
\end{equation}
where $A$ is the adjacency matrix, $M$ is the total number of edges, $k_i$ is the degree of vertex $i$, and $\delta(c_v,c_w) = 1$ if nodes $v$ and $w$ are in the same community and zero otherwise.  Thus $Q$ is the fraction of edges found to be within communities, minus the expected fraction if edges were randomly placed, irrespective of an underlying community structure but respecting degree.  The second term then acts as a null model, and large values of $Q$ indicate deviations away from a random network structure.  

Very efficient algorithms have been created utilizing greedy optimization of $Q$~\cite{newman:fastcommunities1:pre,newmanclausetmoor:fastmodularitypre:2004,1242805}, but any algorithm using $Q$ must necessarily be a global method, requiring complete knowledge of the entire network.  Meanwhile, it has been shown~\cite{resolutionLimitModularity} that $Q$ is not ideal, and a variety of other techniques exist~\cite{citeulike:332503}, but these too generally require global knowledge.  This knowledge isn't available for certain types of networks, such as the WWW, which is simply too large and evolves too quickly to have a fully known structure.  In these circumstances, one must rely on a local method capable of finding a particular community within a network, without knowledge of the structure outside of the discovered community.  Several local methods exist, all of which attempt to find the community containing a particular \emph{starting node}~\cite{flake00efficient,bagrowbollt:lcd,clausetLocalCommunity,DBLP:conf/webi/LuoWP06}.

In this work we present a new technique for quantifying the accuracy of a local method, so that one can determine how various algorithms perform relative to each other.  Due to the unique dependence a local method has upon its starting node, we also develop a simple set of ad hoc benchmark networks, with a generalized degree distribution, allowing one to test accuracy when the starting node is a hub, for example.  We also present a new local method, as well as several types of \emph{stopping criteria} indicating when an algorithm has best found the enclosing community.

\section{Local Community Detection Methods}\label{sec:LCDoverview}
We focus our efforts on two existing algorithms, due to Clauset~\cite{clausetLocalCommunity} and Luo, Wang, and Promislow (LWP)~\cite{DBLP:conf/webi/LuoWP06}, as well as a new method.  Several other local methods exist, including those due to Flake, Lawrence, and Giles~\cite{flake00efficient} and Bagrow and Bollt~\cite{bagrowbollt:lcd}, but these are either reliant on a priori assumptions of network properties (limiting applicability to specific types of networks, such as the WWW), or tend to be accurate only when used as part of a more global method.  Other methods (for example, \cite{Victor-Farutin-1-:2005lr,palla-2005-435,citeulike:151}) concern themselves with local community structure, but either require global knowledge to first determine this structure, or are defined locally but do not provide a definitive partition necessary for evaluation~\cite{CP1,palla-2005-435,CP2,CP3,CP4,CP5,CP6,CP7}.

\begin{figure}
	\centering
	\subfigure[][]{
		\includegraphics[width=0.475\columnwidth,trim=0 25 0 0]{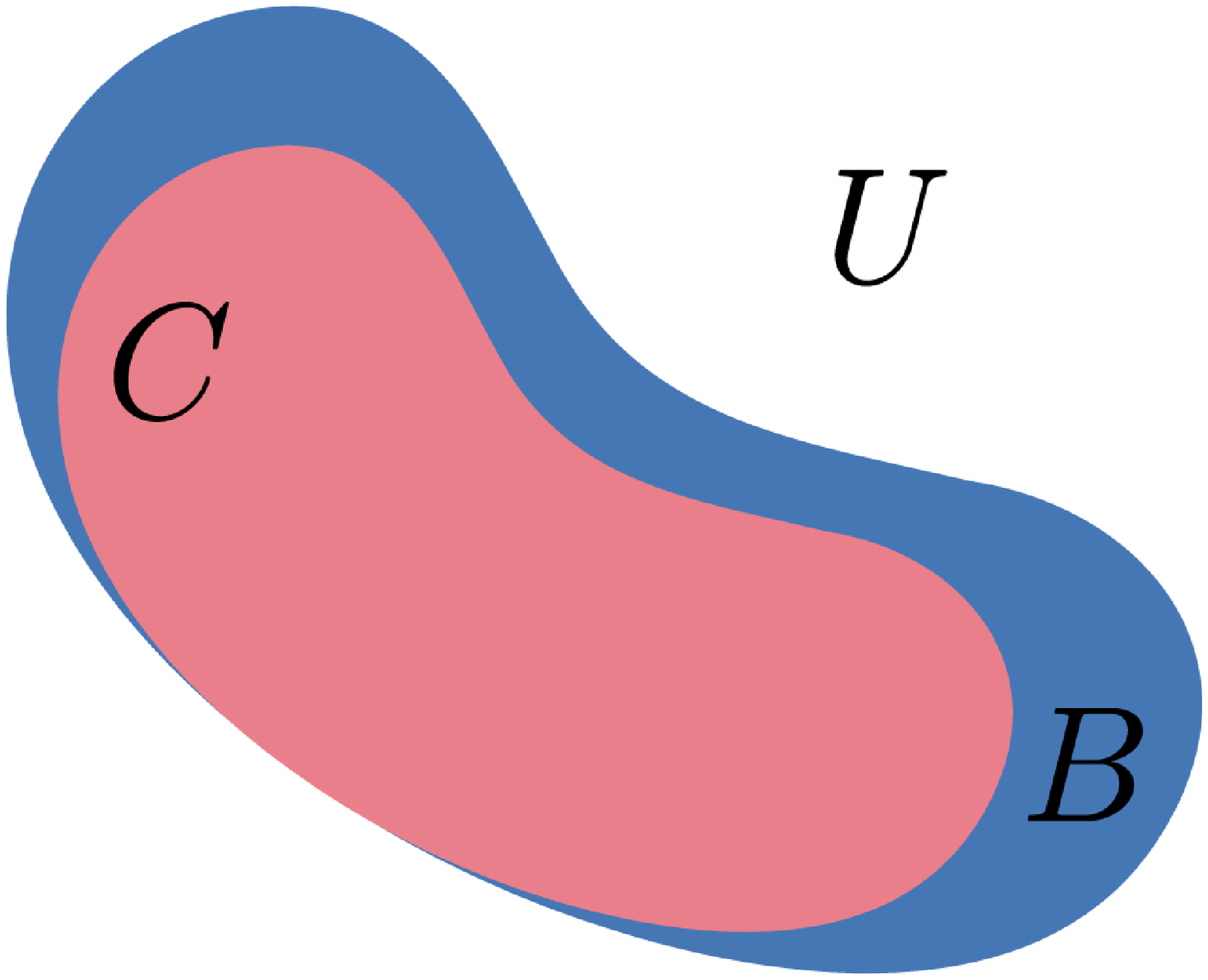}
		\label{subfig:setsDiagram}
		}
	\subfigure[][]{
		\includegraphics[width=0.4\columnwidth]{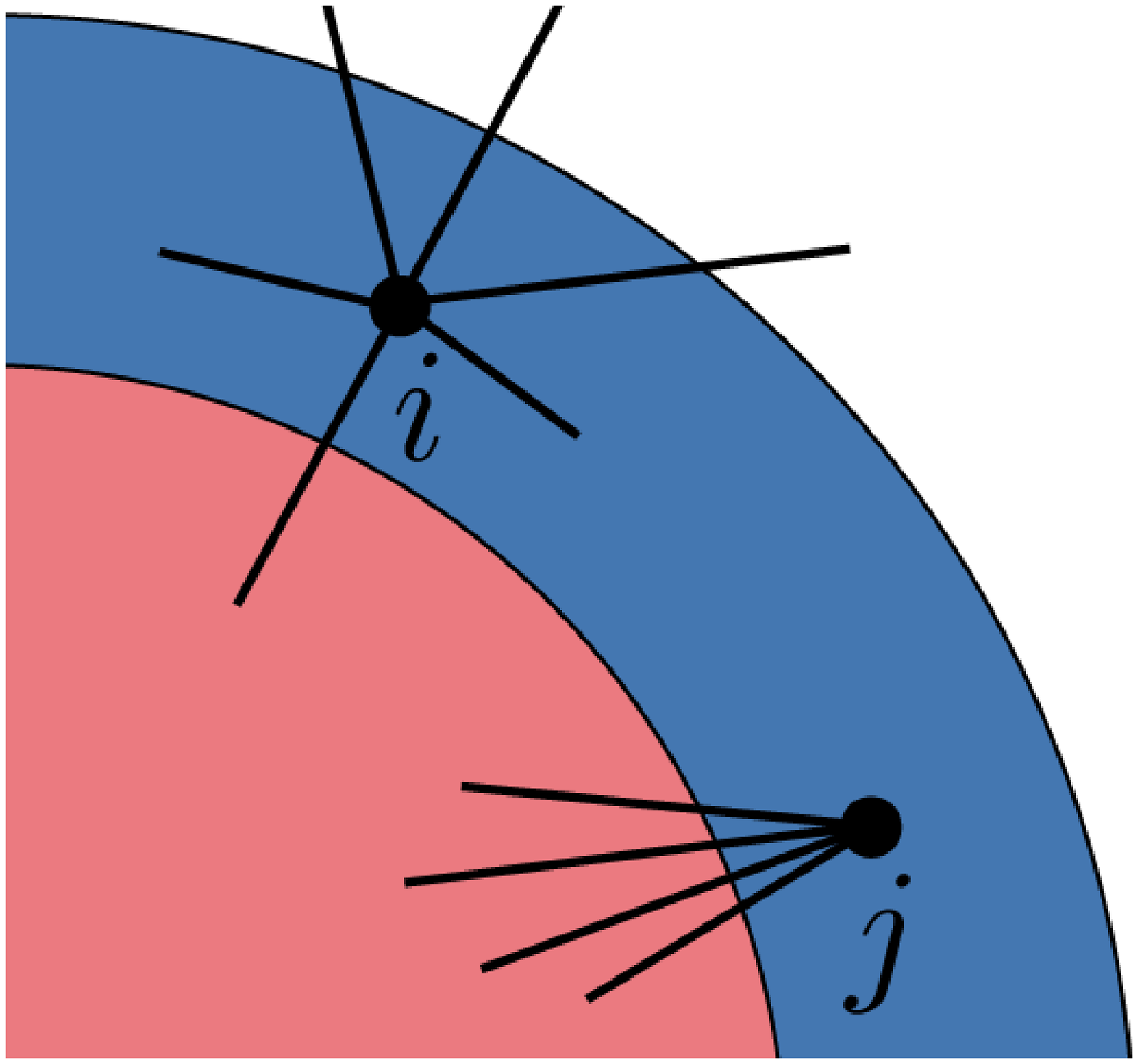}
		\label{subfig:outwardnessDiagram}
	}
	\caption{(color online) (a) The community $C$ is surrounded by a boundary of explored nodes $B$.  This exploration implies an additional layer of nodes that are known only due to their adjacencies with $B$.  (b) Two nodes $i$ and $j$ in $B$, with $\Omega_i = 2/3$ and $\Omega_j = -1$.  Moving node $j$ into $C$ will give improved community structure, compared to moving $i$.}
	\label{fig:outwardness_example}
\end{figure}

All three algorithms begin with a starting node $s$ and divide the explored network into two regions: the community $C$, and the set of nodes adjacent to the community, $B$ (each has at least one neighbor in $C$).  At each step, one or more nodes from $B$ are chosen and agglomerated into $C$, then $B$ is updated to include any newly discovered nodes.  This continues until an appropriate stopping criteria has been satisfied.  When the algorithms begin, $C=\{s\}$ and $B$ contains the neighbors of $s$: $B=\{n(s)\}$.  See Fig.~\ref{subfig:setsDiagram}.

\begin{figure}
	\centering
	\begingroup%
  \makeatletter%
  \newcommand{\GNUPLOTspecial}{%
    \@sanitize\catcode`\%=14\relax\special}%
  \setlength{\unitlength}{0.1bp}%
\begin{picture}(2430,1836)(0,0)%
\includegraphics{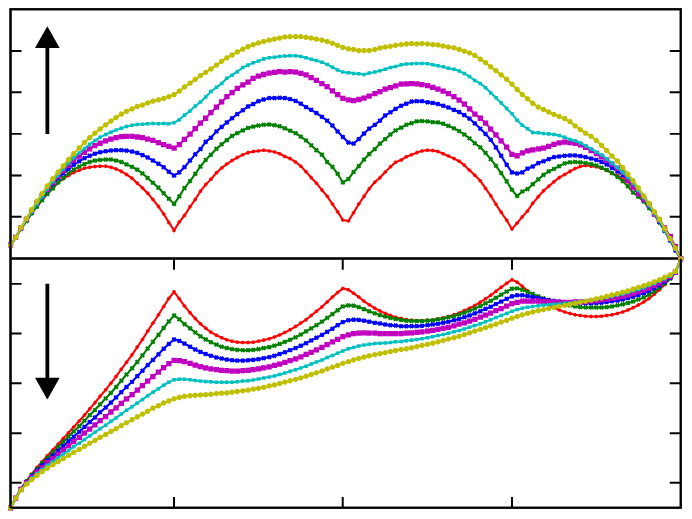}
\put(487,802){\makebox(0,0)[l]{$z_\mathrm{out}$}}%
\put(1315,50){\makebox(0,0){Agglomeration step, community size $|C|$}}%
\put(100,658){%
%
\makebox(0,0)[b]{\shortstack{$R$}}%
%
}%
\put(1794,200){\makebox(0,0){ 96}}%
\put(1307,200){\makebox(0,0){ 64}}%
\put(821,200){\makebox(0,0){ 32}}%
\put(300,945){\makebox(0,0)[r]{ 0.9}}%
\put(300,802){\makebox(0,0)[r]{ 0.7}}%
\put(300,659){\makebox(0,0)[r]{ 0.5}}%
\put(300,515){\makebox(0,0)[r]{ 0.3}}%
\put(300,372){\makebox(0,0)[r]{ 0.1}}%
\put(487,1545){\makebox(0,0)[l]{$z_\mathrm{out}$}}%
\put(100,1377){%
%
\makebox(0,0)[b]{\shortstack{$M_\mathrm{out}$}}%
%
}%
\put(300,1616){\makebox(0,0)[r]{ 250}}%
\put(300,1497){\makebox(0,0)[r]{ 200}}%
\put(300,1377){\makebox(0,0)[r]{ 150}}%
\put(300,1257){\makebox(0,0)[r]{ 100}}%
\put(300,1138){\makebox(0,0)[r]{ 50}}%
\end{picture}%
\endgroup

	\caption{(color online) Comparison between quality measures for the Clauset algorithm, $R$, and the method presented here, $\Mout$.  Shown are the average of 500 realizations of the 128 node ad hoc networks, for $z_\mathrm{out} = 1,2,\ldots,6$. \label{fig:compare_Mout_R_adhoc128} }
\end{figure}

The Clauset algorithm focuses on nodes inside $C$ that form a ``border'' with $B$:  each has at least one neighbor in $B$.  Denoting this set $C_\mathrm{border}$, and focusing on incident edges, Clauset defines the following local modularity:
\begin{equation}
R = \frac{ \sum_{i,j} \beta_{ij} [i \notin B ] [ j \notin B ] } {\sum_{i,j} \beta_{ij} },
\label{eqn:clauset_R}
\end{equation}
where $\beta_{ij}$ is the adjacency matrix comprising only those edges with one or more endpoints in $C_\mathrm{border}$ and $[P]=1$ if proposition $P$ is true, and zero otherwise.  Each node in $B$ that can be agglomerated into $C$ will cause a change in $R$, $\Delta R$, which may be computed efficiently.  At each step, the node with the largest $\Delta R$ is agglomerated. This modularity $R$ lies on the interval $0 \leq R \leq 1$ (defining $R=1$ when $|C_\mathrm{border}|= 0$) and local maxima indicate good community separation, as shown in Fig.~\ref{fig:compare_Mout_R_adhoc128}.  For a network of average degree $d$, the cost to agglomerate $|C|$ nodes is $\mathcal{O}(|C|^2 d)$.  

The LWP algorithm defines a different local modularity, which is closely related to the idea of a \emph{weak} community~\cite{rccvp_defining}.  Define the number of edges internal and external to $C$ as $M_\mathrm{in}$ and $M_\mathrm{out}$, respectively:
\begin{eqnarray}
M_\mathrm{in} &=& \frac{1}{2}\sum_{i,j} A_{ij} [i \in C][j \in C], \\
M_\mathrm{out} &=& \sum_{i,j} A_{ij} [i \in C][j \in B].
\end{eqnarray}
The LWP local modularity $M_f$ is then:
\begin{equation}
M_f(C)  = \frac{M_\mathrm{in}}{M_\mathrm{out}}.
\end{equation}
When $M_f > 1/2$, $C$ is a weak community, according to~\cite{rccvp_defining}.   The algorithm consists of agglomerating \emph{every} node in $B$ that would cause an increase in $M_f$, $\Delta M_f > 0$, then removing every node from $C$ that would also lead to $\Delta M_f > 0$ so long as the node's removal does not disconnect the subgraph induced by $C$. (Removed nodes are not returned to $B$, they are never re-agglomerated.)  Finally $B$ is updated and the process repeats until a step where the net number of agglomerations is zero.  The algorithm returns a community if $M_f > 1$ and $s \in C$.  Similar to the Clauset method, the cost of agglomerating $|C|$ nodes is $\mathcal{O}(|C|^2 d)$.

Finally, we present a new algorithm, as an illustration of how simple an effective local method can be.  Let us define the ``outwardness'' $\Omega_v (C)$ of node $v \in B$ from community $C$:
\begin{eqnarray}
	\Omega_v (C) &=& \frac{1}{k_v} \sum_{i \in n(v)} \Big( \big[i \notin C \big] - \big[i \in C \big] \Big) \\
	             &=& \frac{1}{k_v} \left( k_v^\mathrm{out} - k_v^\mathrm{in} \right)
\end{eqnarray}
where $n(v)$ are the neighbors of $v$.  In other words, the outwardness of a node is the number of neighbors outside the community minus the number inside, normalized by the degree. Thus, $\Omega_v$ has a minimum value of $-1$ if all neighbors of $v$ are inside $C$, and a maximum value of $1 - 2/k_v$, since any $v \in B$ must have at least one neighbor in $C$.  Since finding a community corresponds to maximizing its internal edges while minimizing external ones, we agglomerate the node with the smallest $\Omega$ at each step, breaking ties at random.  See Fig.~\ref{subfig:outwardnessDiagram}.

This method is efficient for the following reasons.  When a node $v \in B$ is moved into $C$, only the neighbors of $v$ will have their outwardness' altered.  For a node $i \in n(v)$, the change in $\Omega_i$ is just $ \Delta\Omega_i = - 2/k_i$ since only a single link can exist between $v$ and $i$.  If node $i$ was not previously in $B$, it will now have a single edge to $C$ and $\Omega_i = 1 - 2/k_i$.  Calculating $\Omega_i$ at each step thus requires knowing only $k_i$, which may be expensive (for example, on the WWW), but needs only be calculated upon the initial discovery of $i$.

For efficiency, one can maintain a min-heap of the outwardness' of all nodes in $B$ then, at each step, extract the minimum with cost $\mathcal{O}(\log|B|)$, and update or insert the neighboring $\Omega$'s.  For a network with average degree $d$, the cost of this updating is $\mathcal{O}(d^2\log|B|)$.  This is often an overestimate, depending on the community structure, since a node's degree need only be calculated once.  Then, the cost of agglomerating $|C|$ nodes is $\mathcal{O}(|C|d^2\log|B|)$.  The relative sizes of $C$ and $B$ are highly dependent on the particular network and the current state of the algorithm, but $|B| \sim |C|$ seems reasonable.  A sparse network with rich community structure would give a cost of $\mathcal{O}(|C|\log|C|)$.

\begin{figure}[!t]
    \centering
    % GNUPLOT: LaTeX picture with Postscript
\begingroup%
  \makeatletter%
  \newcommand{\GNUPLOTspecial}{%
    \@sanitize\catcode`\%=14\relax\special}%
  \setlength{\unitlength}{0.1bp}%
\begin{picture}(2430,1836)(0,0)%
\includegraphics{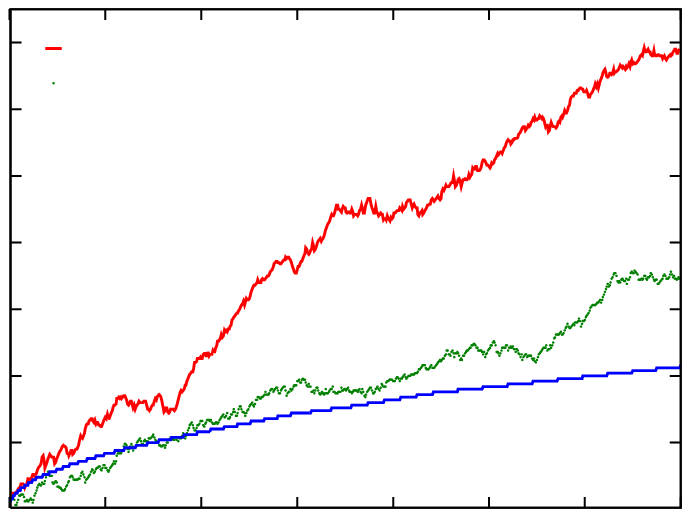}
\put(547,1523){\makebox(0,0)[l]{Planet Earth: BBC Series, DVD}}%
\put(547,1623){\makebox(0,0)[l]{Gravitation: Thorne, Wheeler, Misner}}%
\put(1315,50){\makebox(0,0){Agglomeration step, community size $|C|$}}%
\put(100,1018){%
%
\makebox(0,0)[b]{\shortstack{$M_\mathrm{out}$}}%
%
}%
\put(2280,200){\makebox(0,0){ 700}}%
\put(2004,200){\makebox(0,0){ 600}}%
\put(1728,200){\makebox(0,0){ 500}}%
\put(1452,200){\makebox(0,0){ 400}}%
\put(1176,200){\makebox(0,0){ 300}}%
\put(899,200){\makebox(0,0){ 200}}%
\put(623,200){\makebox(0,0){ 100}}%
\put(347,200){\makebox(0,0){ 0}}%
\put(300,1640){\makebox(0,0)[r]{ 350}}%
\put(300,1448){\makebox(0,0)[r]{ 300}}%
\put(300,1256){\makebox(0,0)[r]{ 250}}%
\put(300,1064){\makebox(0,0)[r]{ 200}}%
\put(300,872){\makebox(0,0)[r]{ 150}}%
\put(300,680){\makebox(0,0)[r]{ 100}}%
\put(300,488){\makebox(0,0)[r]{ 50}}%
\end{picture}%
\endgroup
%\endinput

	\caption{(color online) Comparison of a seminal physics text and a popular DVD (\#1 seller at the time of calculation) on the amazon.com co-purchasing network.  Fluctuations in $\Mout$ in both items indicate the presence of non-trivial community structure. The smooth curve is for a 2D periodic lattice of $500 \times 500$ nodes.\label{fig:amazon_CtoB_Gravitation_PlanetEarth}}
\end{figure}

While seeking to agglomerate the least outward nodes at each step seems natural, it lacks a nicely defined measure of the quality of the community, analogous to $R$ in the Clauset agglomeration.  To overcome this we simply track $\Mout$ during agglomeration.  The smaller this is the better the community separation, so we expect local minima in $\Mout$ when a community has been fully agglomerated.    In addition, $\Mout$  can be easily computed alongside agglomeration.  After agglomerating node $v$, the change in $\Mout$ is just
$\Delta \Mout =  2 k_v^\mathrm{out} - k_v.$  As shown in Fig.~\ref{fig:amazon_CtoB_Gravitation_PlanetEarth}, $\Mout$ provides useful information about a real-world networks' community structure, in this case the amazon.com co-purchasing network~\footnote{This data was generated by crawling the actual links on each amazon product page that point to co-purchased products.  This network evolves over time and results are necessarily altered. }.

Using $\Mout$ as a measure of quality is not ideal, however: it's not normalized, and (like the Clauset modularity) obtains a trivial value when the entire network has been agglomerated.  The latter is less of an issue for local methods.  More worrisome is the fact that $\Mout$ may also be trivially small when $C$ is small.  See Fig.~\ref{fig:compare_Mout_R_adhoc128} for a comparison of $R$ and $\Mout$.  We continue to use $\Mout$ for the sake of simplicity, but more involved measures may certainly lead to improved results.

\section{Stopping Criteria}
After identifying an appropriate agglomeration scheme, a local method must also be able to appropriately \emph{stop} adding nodes.  Here we suggest two possible schemes and will use the techniques and benchmarks of Sec.~\ref{sec:benchmarks} to compare them. It is important that the stopping criteria is also local; a criteria that spreads to the entire network then finds, e.g., the largest values of $\Delta \Mout$ is no longer a local algorithm.

These stopping criteria are essentially divorced from the agglomeration schemes of most local algorithms, allowing one to mix and match to find more accurate methods.  We show this with the Clauset and new method from Sec.~\ref{sec:LCDoverview}.  The LWP algorithm already contains a stopping criteria and we use it unaltered. 

A subgraph $C \subset G$ is a \textbf{strong} community when every node in $C$ has more neighbors inside $C$ than outside \cite{flake00efficient,rccvp_defining}.  This may be used as a local stopping criterion in the following way:  agglomerate nodes until $C$ becomes, and then ceases to be, strong.  Unfortunately, this can be too strict, since a single node can terminate the algorithm.  Define a $p$-strong community as one where this is true for only a fraction $p$ of nodes in $C$.  Then, one can relax the condition by lowering $p$.  Multiple values of $p$ can be used simultaneously, at little cost, and the "best" result (smallest $\Mout > 0$, largest $R < 1$) can be retained as $C$.  We do this for $\{p\}=\{0.75,0.76,\ldots,1\}$.  For specific details, see Appendix~\ref{app:strong}.

Another stopping criterion is what we refer to as Trailing Least-Squares.  Fitting a polynomial to the plot of $\Mout$ during agglomeration, one can identify the cusp or inflection point that indicates a community border.  This method is somewhat involved but our benchmarking procedure shows that it works quite well.  See Appendix~\ref{appendix:TLS}.

\section{Benchmarking}\label{sec:benchmarks}

\subsection{Test graphs}
It has become standard practice to test community algorithms with synthetic networks that possess a given community structure and a parameter to control how well separated the communities are.  The traditional example is the so-called ``ad hoc'' networks~\cite{newman:2004:pre,DDA06}, which typically possesses 128 nodes divided into four equally sized communities.  Each node has (on average) degree $z = z_\mathrm{in} + z_\mathrm{out} = 16$, where $z_\mathrm{out}$ is the number of links a node has to nodes outside its community.  A smaller $z_\mathrm{out}$ (and correspondingly larger $z_\mathrm{in}$) leads to communities that are easier to detect.

These ad hoc networks have a sharply peaked degree distribution.  Since local algorithms are dependent on a particular starting node, their accuracy might be affected if the starting node is a hub or a leaf~\footnote{We term the lowest degree node in the network the ``leaf,'' which is not necessarily of degree 1.}.  So one would also like more realistic synthetic networks which possess a wider degree distribution, such as a power law.  To do this, we propose the following:

\begin{enumerate}
	\item Build a graph $G$ of $N$ nodes and $M$ edges, perhaps using the configuration model and a given degree distribution.  Throughout this work, we use Barab\'asi-Albert graphs of $N = 512$, and $m_0 = 8$~\footnote{These are built quickly by relaxing the constraint on multi-edges, which are then removed~\cite{bb-eglrn-05,networkx}.  The total number of edges will vary slightly, and the lowest degree nodes often have less than $m_0$ neighbors.}.
	
	\item Randomly partition the nodes of $G$ into two or more groups.  These will serve as the ``actual'' communities.  We limit ourselves to four equally sized partitions.
	
	\item Choose random pairs of edges that are \emph{between} the same two groups and rewire them to be \emph{within} the groups, in such a way that the degree distribution is unaltered.
\end{enumerate}
This rewiring (or switching) technique, replacing edges $(i,j)$ and $(k,l)$ with edges $(i,k)$ and $(j,l)$~\cite{2004PhyA..333..529M,citeulike:548}, has been used in the past to \emph{destroy} the presence of community structure, allowing for a null model to test for false positives~\cite{2005PhRvE..71d6101M}.  Here we do the opposite, and communities become more sharply separated as the number of rewirings increases.

Since the partition is random, the initial modularity $Q_0$ will be very small.  As edges are moved within communities, the first sum in Eq.~\eqref{eqn:modularityDefined} will grow but the second term will remain unchanged, since the degree distribution is unaffected.  %Moving a pair of edges from between communities to within will increase $\sum_{v,w}A_{vw} \delta(c_v,c_w)$ by 4, since each edge is counted twice.  
Therefore, the modularity of the actual partition $Q(t)$ after $t$ pairs of edges have been moved is
\begin{equation}
	Q(t) = Q_0 + \frac{2}{M} t.
\end{equation}
Rewiring $M/4$ pairs of edges will give $Q \approx 1/2$, creating an appreciable amount of community structure in the previously randomized graph.

\subsection{Evaluation}\label{subsec:NMI}
Any local method creates a binary partition of the network into the community itself, $C$, and the remaining non-communnity nodes, $\tilde{C} = V - C$.  In a realistic setting $V$ is unknown, but synthetic benchmarks allow one to know the full division.  In addition, for a synthetic benchmark, the \emph{true} partition $P_R = \{C_R, \tilde{C}_R \}$ is already known, while the found partition $P_F = \{C_F,\tilde{C}_F \}$ may differ.

Traditionally, the accuracy of the found communities is quantified by the fraction of correctly identified nodes.  This has been shown to have drawbacks~\cite{DDA06} and the binary partitioning of a local algorithm poses further problems.  For example, if the algorithm fails to stop in time, it has still identified every node in the community correctly, there are just additional nodes incorrectly attributed to that community.  Should each incorrect node give a penalty?  If the algorithm incorrectly finds one community of $N$ nodes, when there were actually $K$ communities of $N/K$ nodes each, one could assign a $+1/N$ for each correct node and $-1/N$ for each incorrect node, giving a composite score of $2/K - 1$.
This means that synthetic networks with different $K$'s cannot be directly compared.  While scores could be subsequently re-normalized to lie between 0 and 1, we propose an alternative that avoids these problems and is unambiguous.

Following the application introduced in~\cite{citeulike:332503}, we use Normalized Mutual Information~\cite{stgh02b,10.1109/CVPR.2003.1211462} to measure how well $P_R$ and $P_F$ correspond to each other:
\begin{equation}
	I(P_R,P_F) = \frac{-2 \sum_i \sum_{j} X_{ij}\log \left(\frac{X_{ij}N}{X_{i.} X_{.j}} \right) }
	{\sum_i X_{i.}\log \left(\frac{X_{i.}}{N}\right) + \sum_j X_{.j} \log\left( \frac{X_{.j}}{N} \right) },
	\label{eqn:NMI_def}
\end{equation}
where $X$ is a $2 \times 2$ matrix with $X_{ij}$ being the number of nodes from real group $i$ that were placed in found group $j$, $X_{.j}=X_{1j}+X_{2j}$, and $X_{i.}=X_{i1}+X_{i2}$.  In a sense, $I(P_R,P_F)$ is a measure of how much is known about partition $P_R$ by knowing partition $P_F$, with $I=1$ corresponding to perfect knowledge, and $I=0$ to no knowledge at all.

In general, the \emph{confusion matrix} $X$ is $N_R \times N_F$ where $N_R$ and $N_F$ are the number of real and found communities, respectively.  The application of Eq.~\eqref{eqn:NMI_def} is a limiting case corresponding to the binary partitioning inherent to local algorithms.

In most figures, we have included a ``faked'' global method, the Clauset-Newman-Moore (CNM) algorithm~\cite{newman:fastcommunities1:pre,newmanclausetmoor:fastmodularitypre:2004}, for comparison.  This was done by running CNM to find the partitioning with the highest modularity, one random community was designated $C$, and the other communities were grouped together in $\tilde{C}$. A local algorithm is unlikely to match the accuracy of a global method, as shown.

\section{Results and Discussion} \label{sec:discussion}
\begin{figure}
      \centering
        \begingroup%
  \makeatletter%
  \newcommand{\GNUPLOTspecial}{%
    \@sanitize\catcode`\%=14\relax\special}%
  \setlength{\unitlength}{0.1bp}%
\begin{picture}(2430,1836)(0,0)%
\includegraphics{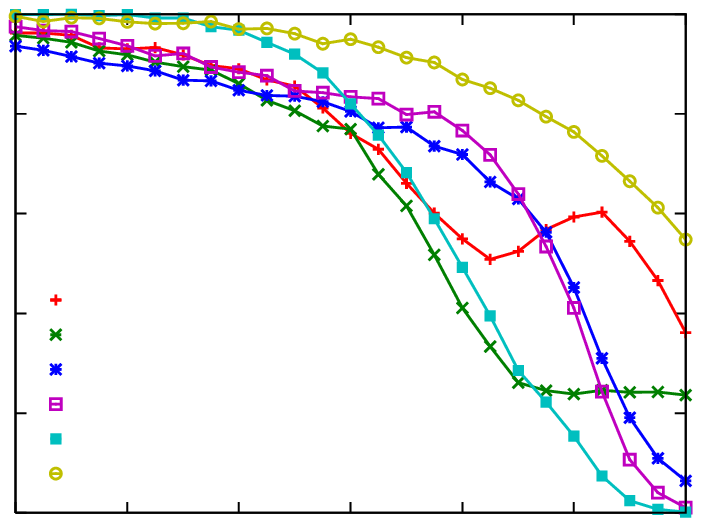}
\put(531,413){\makebox(0,0)[l]{CNM}}%
\put(531,513){\makebox(0,0)[l]{LWP}}%
\put(531,613){\makebox(0,0)[l]{Trailing line, $R$ (Clauset)}}%
\put(531,713){\makebox(0,0)[l]{Trailing parabola, $M_\mathrm{out}$}}%
\put(531,813){\makebox(0,0)[l]{best of $\{p\}$, Clauset}}%
\put(531,913){\makebox(0,0)[l]{best of $\{p\}$}}%
\put(1315,50){\makebox(0,0){$z_\mathrm{out}$}}%
\put(100,1018){%
%
\makebox(0,0)[b]{\shortstack{$I(P_R,P_F)$}}%
%
}%
\put(2280,200){\makebox(0,0){ 7}}%
\put(1958,200){\makebox(0,0){ 6}}%
\put(1637,200){\makebox(0,0){ 5}}%
\put(1315,200){\makebox(0,0){ 4}}%
\put(993,200){\makebox(0,0){ 3}}%
\put(672,200){\makebox(0,0){ 2}}%
\put(350,200){\makebox(0,0){ 1}}%
\put(300,1736){\makebox(0,0)[r]{ 1}}%
\put(300,1449){\makebox(0,0)[r]{ 0.8}}%
\put(300,1162){\makebox(0,0)[r]{ 0.6}}%
\put(300,874){\makebox(0,0)[r]{ 0.4}}%
\put(300,587){\makebox(0,0)[r]{ 0.2}}%
\put(300,300){\makebox(0,0)[r]{ 0}}%
\end{picture}%
\endgroup

	\caption{(color online) An overall comparison of the various methods for the 128-node ad hoc networks, averaged over 1000 realizations.  The LWP method is by far the most accurate for low $z_\mathrm{out}$, while the trailing least-squares methods offer the best performance at higher values.\label{fig:many_orig_bench}}
\end {figure}

\begin {figure}
      \centering
        \begingroup%
  \makeatletter%
  \newcommand{\GNUPLOTspecial}{%
    \@sanitize\catcode`\%=14\relax\special}%
  \setlength{\unitlength}{0.1bp}%
\begin{picture}(2430,1836)(0,0)%
\includegraphics{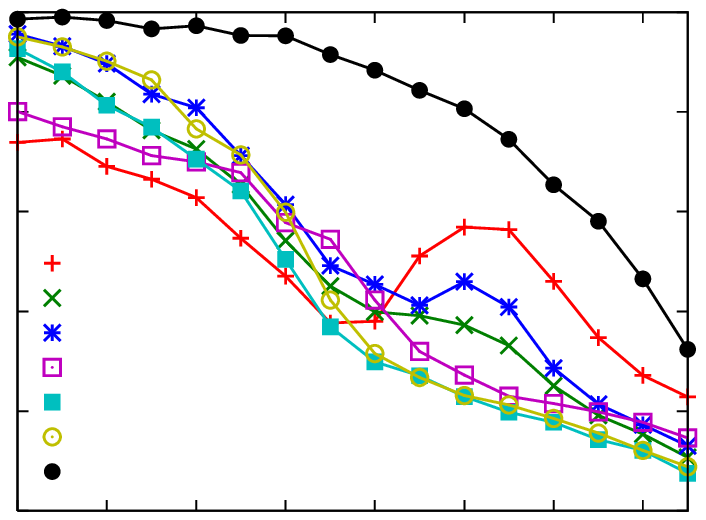}
\put(500,413){\makebox(0,0)[l]{CNM}}%
\put(500,513){\makebox(0,0)[l]{Random, Clauset}}%
\put(500,613){\makebox(0,0)[l]{Leaf, Clauset}}%
\put(500,713){\makebox(0,0)[l]{Hub, Clauset}}%
\put(500,813){\makebox(0,0)[l]{Random}}%
\put(500,913){\makebox(0,0)[l]{Leaf}}%
\put(500,1013){\makebox(0,0)[l]{Hub}}%
\put(1315,50){\makebox(0,0){number of rewirings, $t$}}%
\put(100,1018){%
%
\makebox(0,0)[b]{\shortstack{$I(P_R,P_F)$}}%
%
}%
\put(350,200){\makebox(0,0){ 1300}}%
\put(607,200){\makebox(0,0){ 1200}}%
\put(865,200){\makebox(0,0){ 1100}}%
\put(1122,200){\makebox(0,0){ 1000}}%
\put(1379,200){\makebox(0,0){ 900}}%
\put(1637,200){\makebox(0,0){ 800}}%
\put(1894,200){\makebox(0,0){ 700}}%
\put(2151,200){\makebox(0,0){ 600}}%
\put(300,1736){\makebox(0,0)[r]{ 1}}%
\put(300,1449){\makebox(0,0)[r]{ 0.8}}%
\put(300,1162){\makebox(0,0)[r]{ 0.6}}%
\put(300,874){\makebox(0,0)[r]{ 0.4}}%
\put(300,587){\makebox(0,0)[r]{ 0.2}}%
\put(300,300){\makebox(0,0)[r]{ 0}}%
\end{picture}%
\endgroup

\caption{(color online) Using the ``best of $\{p\}$-strong'' criteria on the 512-node rewired networks, for $\{p\}=0.75,0.76,\ldots,1$.  Each point averaged over 500 realizations.  The effect of rejecting any individual p-strong results where $\Mout = 0$ $(R = 1)$ is more apparent for these networks, especially for hub nodes.\label{fig:rewireManyP}}
\end {figure}

\begin {figure}
      \centering
        \begingroup%
  \makeatletter%
  \newcommand{\GNUPLOTspecial}{%
    \@sanitize\catcode`\%=14\relax\special}%
  \setlength{\unitlength}{0.1bp}%
\begin{picture}(2430,1836)(0,0)%
\includegraphics{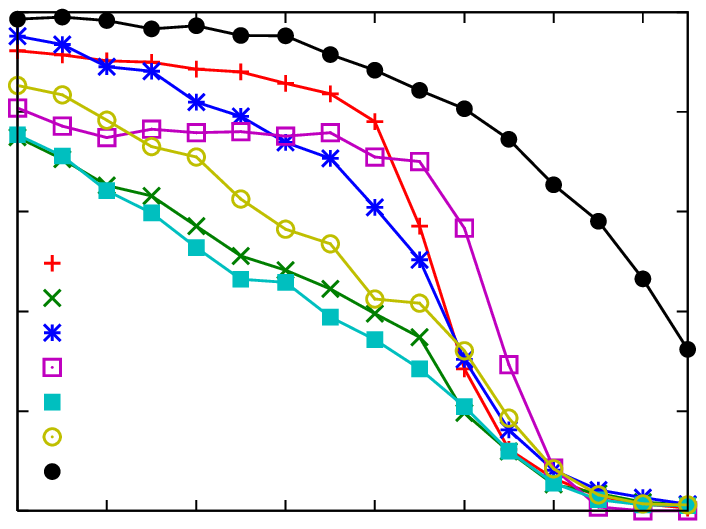}
\put(500,413){\makebox(0,0)[l]{CNM}}%
\put(500,513){\makebox(0,0)[l]{Random, Clauset}}%
\put(500,613){\makebox(0,0)[l]{Leaf, Clauset}}%
\put(500,713){\makebox(0,0)[l]{Hub, Clauset}}%
\put(500,813){\makebox(0,0)[l]{Random}}%
\put(500,913){\makebox(0,0)[l]{Leaf}}%
\put(500,1013){\makebox(0,0)[l]{Hub}}%
\put(1315,50){\makebox(0,0){number of rewirings, $t$}}%
\put(100,1018){%
%
\makebox(0,0)[b]{\shortstack{$I(P_R,P_F)$}}%
%
}%
\put(350,200){\makebox(0,0){ 1300}}%
\put(607,200){\makebox(0,0){ 1200}}%
\put(865,200){\makebox(0,0){ 1100}}%
\put(1122,200){\makebox(0,0){ 1000}}%
\put(1379,200){\makebox(0,0){ 900}}%
\put(1637,200){\makebox(0,0){ 800}}%
\put(1894,200){\makebox(0,0){ 700}}%
\put(2151,200){\makebox(0,0){ 600}}%
\put(300,1736){\makebox(0,0)[r]{ 1}}%
\put(300,1449){\makebox(0,0)[r]{ 0.8}}%
\put(300,1162){\makebox(0,0)[r]{ 0.6}}%
\put(300,874){\makebox(0,0)[r]{ 0.4}}%
\put(300,587){\makebox(0,0)[r]{ 0.2}}%
\put(300,300){\makebox(0,0)[r]{ 0}}%
\end{picture}%
\endgroup

\caption{\label{fig:TrailingRewire} (color online) A comparison of the trailing least-squares criteria for both the new algorithm and the Clauset method.  Starting from a hub tends to be the most accurate, except when the communities are very well separated.}
\end {figure}

\begin {figure}
      \centering
        \begingroup%
  \makeatletter%
  \newcommand{\GNUPLOTspecial}{%
    \@sanitize\catcode`\%=14\relax\special}%
  \setlength{\unitlength}{0.1bp}%
\begin{picture}(2430,1836)(0,0)%
\includegraphics{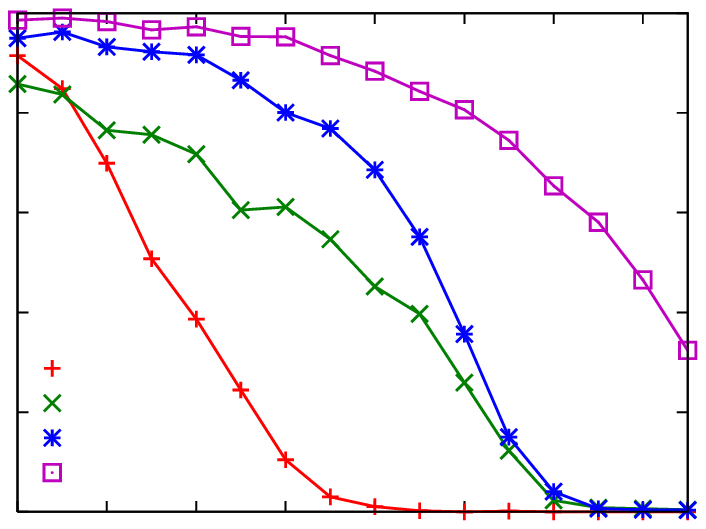}
\put(500,413){\makebox(0,0)[l]{CNM}}%
\put(500,513){\makebox(0,0)[l]{Random}}%
\put(500,613){\makebox(0,0)[l]{Leaf}}%
\put(500,713){\makebox(0,0)[l]{Hub}}%
\put(1315,50){\makebox(0,0){number of rewirings, $t$}}%
\put(100,1018){%
%
\makebox(0,0)[b]{\shortstack{$I(P_R,P_F)$}}%
%
}%
\put(350,200){\makebox(0,0){ 1300}}%
\put(607,200){\makebox(0,0){ 1200}}%
\put(865,200){\makebox(0,0){ 1100}}%
\put(1122,200){\makebox(0,0){ 1000}}%
\put(1379,200){\makebox(0,0){ 900}}%
\put(1637,200){\makebox(0,0){ 800}}%
\put(1894,200){\makebox(0,0){ 700}}%
\put(2151,200){\makebox(0,0){ 600}}%
\put(300,1736){\makebox(0,0)[r]{ 1}}%
\put(300,1449){\makebox(0,0)[r]{ 0.8}}%
\put(300,1162){\makebox(0,0)[r]{ 0.6}}%
\put(300,874){\makebox(0,0)[r]{ 0.4}}%
\put(300,587){\makebox(0,0)[r]{ 0.2}}%
\put(300,300){\makebox(0,0)[r]{ 0}}%
\end{picture}%
\endgroup

\caption{\label{fig:LWPrewired}  (color online)  The LWP algorithm used on the rewired benchmark networks.  LWP performs very well for large numbers of rewirings, but becomes progressively worse as less edges are moved. Both extremes, hubs and leaves, decrease overall accuracy.}
\end {figure}

The results of simulations, shown in Figs.~\ref{fig:many_orig_bench}--\ref{fig:LWPrewired}, indicate the relative accuracies of the various algorithms and stopping criteria.  As shown in Figs. \ref{fig:many_orig_bench} and \ref{fig:LWPrewired}, the LWP method performs extremely well for clearly separated communities, with a rapid decrease in accuracy as the separation blurs.  %LWP has the additional benefit of being the only algorithm to have essentially zero dependence on the degree of the starting node, as shown in Fig.~\ref{fig:LWPrewired}.

The ``best of $\{p\}$-strong'' (Figs. 6 and 7) and trailing least-squares (Figs. 6 and 8) stopping criteria first perform at comparable accuracy for both algorithms for the 128-node ad hoc networks, but the trailing least-squares tends to perform better as community distinction blurs.  Trailing least-squares outperforms $\{p\}$-strong in the 512-node networks (Fig. 8 vs. Fig. 9), suggesting that the size of the community impacts accuracy (which might be expected when fitting data).  

Overall, the best of $\{p\}$-strong has the least accuracy but is also least affected by the degree of the starting node.  Meanwhile, trailing least-squares performs better overall but is more dependent on the starting node.  The LWP algorithm is also quite accurate overall, though trailing least-squares can outperform it when the community separation is less clear.

%\section{Results}

The agglomeration schemes presented share many similarities, and a certain amount of ``cross-pollination'' is possible.  For example, accuracy may improve if one maintains the outwardness of nodes after agglomeration and, as per LWP, remove every node from $C$ with positive outwardness.  Another possibility is simply agglomerating all nodes with the minimum $\Omega$ together, instead of breaking ties.  This is not necessarily a trivial difference: the agglomeration histories may diverge since the sequence of nodes exposed to $B$ can differ.

There is much room open to develop accurate stopping criteria.  For example, the notion of a weak community can also be generalized to provide a (perhaps improved) stopping criteria.  As defined, a community is weak when $M_\mathrm{in} > \frac{1}{2} \Mout$.  This can be generalized by introducing a parameter to control how strict the constraint is:  a community is $p$-weak when $M_\mathrm{in} > p \Mout$.  Thus, a weak community corresponds to $\frac{1}{2}$-weak, and the LWP stopping criteria is $1$-weak.   While the introduction of a further parameter is not ideal, and the lack of performance of the $p$-strong criteria versus the trailing least-squares is not promising, it may still be worth pursuing this and other, similar stopping criteria.  Furthermore, stopping criteria using $LS$-sets and $k$-cores, as mentioned in~\cite{rccvp_defining}, may also be worth investigation.

In addition to finding a single community, any local method could be easily adapted to find more community structure, simply by running the local algorithm multiple times (possibly without repeated agglomeration of nodes or similar modifications).  These \emph{quasi-local} methods may not have the same level of accuracy as a global method --- agglomerating communities sequentially may lead to compounding errors --- but it may still be worth pursuing, even if only as an initialization step for a different algorithm.  

There is an implicit assumption, in all these methods, that the underlying network is truly undirected.  Of course, this is not generally true.  In the WWW it is easy to know what pages an explored web page links to, but it is impossible to know how many other pages may link to the explored page.  These \emph{back links} are simply disregarded by the local methods, and it seems a difficult problem to overcome, especially when applying a quasi-local method and back links continue to be discovered as more communities are found.  One possible way to overcome this is to maintain $\Omega_v$ after agglomeration, then go through all the found communities, remove nodes with, say, $\Omega > 0$, then re-agglomerate them into the community with the smallest outwardness.  Another idea, suggested in~\cite{flake00efficient} is to use a global index, such as a search engine, to list all the back links. It seems that in a different context, such as a partially explored social network, one has no choice but to ignore these back links until they are discovered, then adjust the results accordingly.

\section{Conclusions}
Much recent work has been applied to the problem of finding communities in complex networks.  In this paper, we have focused on the idea of finding a particular community inside of a network without relying on global knowledge of the entire network's structure, knowledge that is unavailable in a variety of areas.  We have introduced a new and very simple local method, with a running time of $\mathcal{O}(|C| \log |C|)$.  Several types of stopping criteria have been introduced, which can be used in conjunction with different agglomeration schemes.  

Using Normalized Mutual Information, we have introduced a simple and unambiguous means of quantifying the accuracy of a local algorithm when applied to a synthetic network with pre-defined community structure.  Synthetic networks with generalized degree distributions have been used to allow one to test the impact of the starting node's degree, something not possible with existing ad hoc networks.  

These techniques have been applied to compare the accuracy of a variety of agglomeration schemes and stopping criteria and we feel they will be of great use when testing newly designed local algorithms.  The fact that multiple stopping criteria and algorithms can perform with comparable accuracy shows that 
the community problem is ill-posed to the point of requiring heuristic
methods, and thus it is worth using an evaluation scheme to compare and
contrast alternative methods.

\appendix
\section{Strong Communities}\label{app:strong}
As per~\cite{flake00efficient,rccvp_defining}, a subgraph $C \subset G$ is a \textbf{strong} community (denoted ``ideal'' in~\cite{flake00efficient}) when every node in $C$ has more neighbors inside $C$ than outside:
\begin{equation}
 k_{i}^{\mathrm{in}}(C) > k_{i}^{\mathrm{out}}(C) , \;\; \forall i \in C.
\label{eqn:strong_comm_radicchi}
\end{equation}
This local quantity allows for a very simple, natural stopping criteria: agglomerate nodes until the community becomes strong then, at each agglomeration step, check $k^{\mathrm{in}}$ and $k^{\mathrm{out}}$ for the newly chosen node and stop agglomerating if the community would cease to be strong.  If $C$ never becomes strong, the algorithm won't terminate, indicating a possible lack of community structure in the explored region of the network.

\begin{figure}[t]
	\centering
        \begingroup%
  \makeatletter%
  \newcommand{\GNUPLOTspecial}{%
    \@sanitize\catcode`\%=14\relax\special}%
  \setlength{\unitlength}{0.1bp}%
\begin{picture}(2430,1836)(0,0)%
\includegraphics{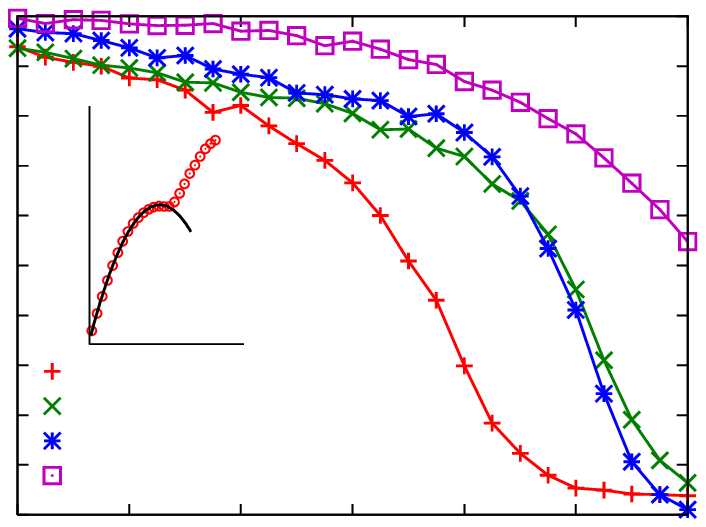}
\put(779,866){\makebox(0,0){$|C|$}}%
\put(532,1134){%
%
\makebox(0,0)[b]{\shortstack{$M_\mathrm{out}$}}%
%
}%
\put(500,413){\makebox(0,0)[l]{CNM}}%
\put(500,513){\makebox(0,0)[l]{Trailing line, $R$ (Clauset)}}%
\put(500,613){\makebox(0,0)[l]{Trailing parabola, $M_\mathrm{out}$}}%
\put(500,713){\makebox(0,0)[l]{Strong to not}}%
\put(1315,50){\makebox(0,0){$z_\mathrm{out}$}}%
\put(100,1018){%
%
\makebox(0,0)[b]{\shortstack{$I(P_R,P_F)$}}%
%
}%
\put(2280,200){\makebox(0,0){ 7}}%
\put(1958,200){\makebox(0,0){ 6}}%
\put(1637,200){\makebox(0,0){ 5}}%
\put(1315,200){\makebox(0,0){ 4}}%
\put(993,200){\makebox(0,0){ 3}}%
\put(672,200){\makebox(0,0){ 2}}%
\put(350,200){\makebox(0,0){ 1}}%
\put(300,1736){\makebox(0,0)[r]{ 1}}%
\put(300,1592){\makebox(0,0)[r]{ 0.9}}%
\put(300,1449){\makebox(0,0)[r]{ 0.8}}%
\put(300,1305){\makebox(0,0)[r]{ 0.7}}%
\put(300,1162){\makebox(0,0)[r]{ 0.6}}%
\put(300,1018){\makebox(0,0)[r]{ 0.5}}%
\put(300,874){\makebox(0,0)[r]{ 0.4}}%
\put(300,731){\makebox(0,0)[r]{ 0.3}}%
\put(300,587){\makebox(0,0)[r]{ 0.2}}%
\put(300,444){\makebox(0,0)[r]{ 0.1}}%
\put(300,300){\makebox(0,0)[r]{ 0}}%
\end{picture}%
\endgroup

	\caption{(color online) The ``strong to not''  and trailing least-squares stopping criteria for the 128-node ad hoc networks using the Clauset method and the new algorithm presented here. Each point is averaged over 1000 realizations.\label{fig:plot_trailingParabola_vs_Z_runs500}}
\end{figure}

As shown in Fig.~\ref{fig:plot_trailingParabola_vs_Z_runs500}, this ``strong to not'' criteria works well for sharply separated communities, but tends to fail as  the contrast decreases.  In a sense, a strong community is \emph{too} strong of a requirement: as the distinction between communities blurs, some nodes must fail Eq.~\eqref{eqn:strong_comm_radicchi}, despite probable membership in $C$.

\begin{figure}
      \centering
        \begingroup%
  \makeatletter%
  \newcommand{\GNUPLOTspecial}{%
    \@sanitize\catcode`\%=14\relax\special}%
  \setlength{\unitlength}{0.1bp}%
\begin{picture}(2430,1836)(0,0)%
\includegraphics{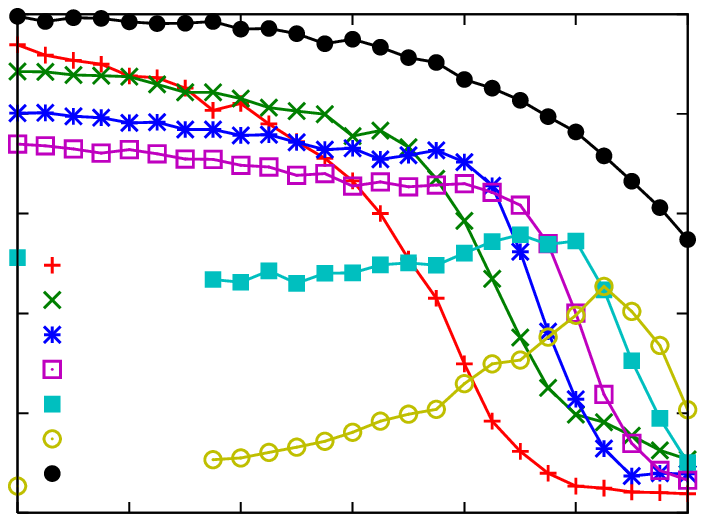}
\put(500,413){\makebox(0,0)[l]{CNM}}%
\put(500,513){\makebox(0,0)[l]{$p=0.75$}}%
\put(500,613){\makebox(0,0)[l]{$p=0.8$}}%
\put(500,713){\makebox(0,0)[l]{$p=0.85$}}%
\put(500,813){\makebox(0,0)[l]{$p=0.9$}}%
\put(500,913){\makebox(0,0)[l]{$p=0.95$}}%
\put(500,1013){\makebox(0,0)[l]{$p=1.0$}}%
\put(1315,50){\makebox(0,0){$z_{\mathrm{out}}$}}%
\put(100,1018){%
%
\makebox(0,0)[b]{\shortstack{$I(P_R,P_F)$}}%
%
}%
\put(2280,200){\makebox(0,0){ 7}}%
\put(1958,200){\makebox(0,0){ 6}}%
\put(1637,200){\makebox(0,0){ 5}}%
\put(1315,200){\makebox(0,0){ 4}}%
\put(993,200){\makebox(0,0){ 3}}%
\put(672,200){\makebox(0,0){ 2}}%
\put(350,200){\makebox(0,0){ 1}}%
\put(300,1736){\makebox(0,0)[r]{ 1}}%
\put(300,1449){\makebox(0,0)[r]{ 0.8}}%
\put(300,1162){\makebox(0,0)[r]{ 0.6}}%
\put(300,874){\makebox(0,0)[r]{ 0.4}}%
\put(300,587){\makebox(0,0)[r]{ 0.2}}%
\put(300,300){\makebox(0,0)[r]{ 0}}%
\end{picture}%
\endgroup

	\caption{(color online) Comparison of various $p$-strong stopping criteria for the 128 node ad hoc networks using the new algorithm shown in Sec.~\ref{sec:LCDoverview}.\label{fig:plot_each_pStrong_vs_Z_runs1000}}
\end{figure}

We generalize the notion of a strong community in the following way.  A community is $p$-\textbf{strong} if Eq.~\eqref{eqn:strong_comm_radicchi} holds, not for all, but only a fraction $p$ (or more) of the nodes:
\begin{equation}
	\sum_{i \in C} \Big[ k_{i}^{\mathrm{in}}(C) > k_{i}^{\mathrm{out}}(C) \Big] \geq p \left| C \right|.
	\label{eqn:pstrong}
\end{equation}
Equations~\eqref{eqn:strong_comm_radicchi} and \eqref{eqn:pstrong} are equivalent when $p=1$, while the requirement becomes increasingly lenient as $p$ decreases.  This allows one to tune the sensitivity by varying $p$.   See Fig.~\ref{fig:plot_each_pStrong_vs_Z_runs1000}.

An additional benefit of Eq.~\eqref{eqn:pstrong} is that multiple values of $p$ can be used simultaneously~\footnote{Indeed, since stopping criteria are often divorced from agglomeration, all manner of criteria may be used simultaneously, to the point where testing to stop can be more expensive than agglomerating.
}, since a community that is $p_1$-strong is also $p_2$-strong ($p_1 > p_2$).  More specifically, for the actual fraction $p_\mathrm{eff}$,
\begin{equation}
	p_\mathrm{eff} = 	\frac{1}{\left| C \right|} \sum_{i \in C} \Big[ k_{i}^{\mathrm{in}}(C) > k_{i}^{\mathrm{out}}(C) \Big],
	\label{eqn:peff}
\end{equation}
$C$ is $p$-strong for all $p \leq p_\mathrm{eff}$, and not $p$-strong for all $p>p_\mathrm{eff}$. 

To use, simply choose a set of appropriate parameters, $\{ p_1,p_2,\ldots\}$, perform the local algorithm, and maintain the state of $C$ as each $p_i$ stopping criteria is satisfied.  One can further use a quality value, such as $\Mout$ or $R$, and choose the best corresponding $C_i$ (in this case, that with the smallest $\Mout$ or largest $R$~\footnote{We limit ourselves to choosing the smallest $\Mout > 0$ ($R < 1$), unless \textit{every} $C_i$ has $\Mout = 0$ ($R=1$).  This distinction is important for finite graphs, causing a curious (and artificial) increase in accuracy for larger values of $z_\mathrm{out}$ (smaller numbers of rewirings).  This is because inaccurate results that previously spread to \emph{most} of the network now spread to the \emph{entire} network and are subsequently being ignored, raising the average value of $I(P_R,P_F)$.}).  This ``best of $\{p\}$'' stopping criterion does not entirely negate the introduction of a new parameter; choosing $p$ too small (e.g. $p=0.1$) can lead to stopping very early.  For this work, we use $\{p\} = \{0.75,0.76,\ldots,1.0\}$, but this may be worth further exploration.  See Figs.~\ref{fig:many_orig_bench} and \ref{fig:rewireManyP}.

%While one may expect the best choice of a set of $p$'s to always perform better than a single appropriately chosen $p$, this is not necessarily true.  This is because every $p_i$'s corresponding $C_i$ is found using the same starting node which, if poorly positioned, will affect all the results.  This is shown in Fig.~\ref{fig:compare_varyingStartingNodes} by comparing the best results from $\{p\}$ all beginning from the same starting node versus each value starting from a randomly chosen starting node.  The latter performs much better, since it's really performing a non-local sampling of the network.

%For the $\{p\}$-strong stopping criteria, choosing the ``best'' community consists of choosing the result with the smallest $\Mout$ (or largest $R$ in the case of the Clauset method).   Since these synthetic graphs are finite, one can spread to the entire network and reach the trivial value.  To overcome this, we instead choose the result with the smallest $\Mout > 0$ ($R<1$).    ***

In addition to strong communities, \emph{weak} communities have been defined~\cite{rccvp_defining}.  A community is weak when $M_\mathrm{in} > \frac{1}{2} M_\mathrm{out}$.  We have found the usage of a ``weak-to-not'' stopping criteria to be problematic.  The impact of a single agglomeration is so small that the community will blissfully continue to grow, far past the appropriate stopping point.  Just as the strong stopping criteria is too strong, a weak stopping criteria is too weak. See Sec.~\ref{sec:discussion} for further ideas, however.

\section{Trailing Least-Squares}\label{appendix:TLS}
Inspired by plots of $R$ and $\Mout$, and in an effort to increase accuracy when community structure is less favorable, we propose another stopping criteria, based on fitting a polynomial to $\Mout$ (or $R$) to find local minima/maxima. Suppose $n$ nodes have been agglomerated, fit $y=a x^2 + b x + c$ to the first $n-3$ values of $\Mout$.  Then extrapolate $y$ to points $n-2$, $n-1$, $n$ and test the following:
\begin{enumerate}
	\item parabola opens downward, $a < 0$ \textbf{and},
	\item $\Mout(i) > y(i)$, $i=n, n-1, n-2$ \textbf{and},
	\item $n-3 > -b/2a$ \textbf{and},
	\item $\Mout(n) \geq \Mout(n-1) \geq \Mout(n-2)$.
\end{enumerate}
If all are satisfied, stop agglomerating (and remove the final three nodes). 

As shown in Fig.~\ref{fig:plot_trailingParabola_vs_Z_runs500}'s inset, when you pass the border of the community, $\Mout$ will start to increase, while the parabola, unaware of the next three values, continues downward.  This works whether the minima is a cusp or just an inflection point, so one need not resort to testing first versus second differences in $\Mout$, etc.  The fitting also provides a degree of smoothing.

This criteria is somewhat involved and has several semi-arbitrary factors: one could extrapolate to a different number of points, relax some of the constraints, fit a different order polynomial, continue fitting until the criteria ceases to be satisfied, etc.  Our results indicate that this criteria as chosen works well, but further refinement is certainly possible. We also use this criteria by fitting a line to $R$ from the Clauset method, since Eq.~\eqref{eqn:clauset_R} tends to grow linearly in the first community.  Both fits have similar accuracy, as shown in Fig.~\ref{fig:plot_trailingParabola_vs_Z_runs500}.

\acknowledgments
We thank E.~Bollt, D.~ben-Avraham, and especially H.~Rozenfeld for useful discussions; A.~Clauset for discussions and shared source code; and A.~Harkin, W.~Basener, and the RIT math department for their hospitality and feedback.  This material is based upon work supported under a National Science Foundation Graduate Research Fellowship.

\end{document}